\documentclass[twocolumn]{aa}
\usepackage{graphicx}
\usepackage{natbib}
\bibpunct{(}{)}{;}{a}{}{,}
\usepackage{txfonts}
\begin{document}
   \title{Globular cluster candidates within the Fornax Cluster: 
   intracluster globulars? \thanks{Based on observations obtained at 
   Cerro Tololo Inter-American Observatory, NOAO, which is operated
   by AURA, Inc., under cooperative agreement with the National 
   Science Foundation.}}

   \author{L. P. Bassino \inst{1,2}
          \and
          S. A. Cellone \inst{1}
          \and
          J. C. Forte \inst{1,2}
          \and
          B. Dirsch \inst{3} 
          }

   \offprints{L. P. Bassino, \\
   \email{lbassino@fcaglp.fcaglp.unlp.edu.ar}}

   \institute{Facultad de Ciencias Astron\'omicas y Geof\'{\i}sicas, 
   Universidad Nacional de La Plata, Paseo del Bosque S/N, 
   1900--La Plata, Argentina
         \and
             Consejo Nacional de Investigaciones Cient\'{\i}ficas y 
   T\'ecnicas (CONICET), Argentina.
         \and
             Universidad de Concepci\'on, Departamento de F\'{\i}sica,
   Casilla 160-C, Concepci\'on, Chile 
             }

   \date{Received date ; Accepted date}

   \abstract{
   We present the results of a search for globular clusters in
   the surroundings of 15 low surface brightness dwarf galaxies belonging
   to the Fornax Cluster, which was carried out on CCD images obtained with
   the $C$ and $T_1$ filters of the Washington photometric system.
   The globular cluster candidates show an extended and probably 
   bimodal $(C-T_1)$ color
   distribution, which is inconsistent with the presence of a single 
   population of metal-poor clusters detected in several dwarf galaxies.
   The surface number density of these candidates shows no
   concentration towards the respective dwarf galaxies, in whose outskirts
   they have been identified.
   On the contrary, if we split the candidates in two groups according  
   to their projected distances to the center of the Fornax Cluster, those 
   located closer to the center show a higher projected density 
   than those located farther from it. These results suggest that 
   the potential globular clusters might not be bound to the dwarf 
   galaxies. Alternatively, these globulars could form part of the very 
   peripheral regions of \object{NGC 1399} (the central galaxy of the Fornax
   Cluster) or even belong to the intracluster medium.

   }

   \titlerunning{Globular cluster candidates in Fornax}
   \authorrunning{Bassino, Cellone, Forte \& Dirsch}

   \maketitle

\section{Introduction}
    
   The Fornax Cluster is an excellent target for
   studying globular clusters: it is very rich, contains
   different types of galaxies and the globular cluster
   candidates can be detected at least as unresolved sources.
   There are numerous photometric studies of globular cluster systems  
   in selected Fornax galaxies, particularly the central one 
   \object{NGC 1399} as well as other early-type galaxies. 
   Most of these studies 
   show that the color distribution of the globular clusters is bimodal 
    due to the presence of two globular cluster populations,
   ``red" and ``blue"; these integrated colors are mainly driven by
   metallicity in objects as old as these ones.
   
   With regard to the globular cluster system of \object{NGC 1399}, the 
   Washington photometry by \citet{ost93} confirmed the existence 
   of a color gradient that had been detected earlier by \citet{bri91}, and 
   suggested that the color distribution was multimodal, as  
   was also supported by the V,I photometry by \citet{kis97}. 
   The HST imaging by \citet{for98}
   and \citet{gri99}, and a refined analysis of their previous Washington 
   photometry by \citet{ost98}, pointed to the bimodal character of the 
   color distribution. More recently, the wide-field 
   study by \citet{di02a,di02b} showed that globular custers with a broad
   metallicity distribution -- that cannot be fitted with a single 
   Gaussian -- extend further than 100 kpc from \object{NGC 1399} center.

   There are fewer investigations of globular cluster systems in dwarf 
   galaxies that are not in the Local Group. Miller and collaborators 
   \citep{mi98a,mi98b,mil99,lot01} carried out
   a survey with images from the Wide Field Planetary
   Camera 2 of the Hubble Space Telescope (FOV $\approx$ 6~arcmin$^2$)
   to analyze the properties of globular clusters and nuclei
   of dwarf elliptical galaxies (dEs) in the Fornax and Virgo Clusters
   and the Leo Group. They include about 20 dEs from the Fornax Cluster
   but none of them are in common with the present work. 
   They show that the globular cluster specific frequency $S_\mathrm{N}$ 
   (number of globular clusters with respect 
   to the parent galaxy's luminosity) of the dEs is in the range 2--6;
   the luminosity function  of  the globular cluster candidates is  consistent
   with a  Gaussian with peak at $M_V^0 \approx - 7$ mag and dispersion
   $\sigma_V \approx~1.4$ mag (assuming a distance modulus of 31.4 for 
   the Fornax Cluster); and most of the globular cluster $(V-I)$ colors are 
   similar to those of the metal-poor globular cluster population
   \citep[also][]{ash93}. 
   The globular cluster system of the luminous dE,N galaxy 
   \object{NGC 3115 DW1} was studied by \citet{du96a} and \citet{puz00} 
   who obtained  mean metallicities $[Fe/H]= -1.2$ and $-1$, respectively,
   and they both agreed on a specific frequency $S_\mathrm{N}$ = 4.9. 
   \citet{du96b} obtained, on the basis of Washington photometry, a 
   low mean metallicity, $[Fe/H]= -1.45$, for the globular cluster systems 
   of two dE galaxies in the Virgo Cluster and they suggested that the dwarf 
   galaxies globular cluster systems present a similar range of metallicities 
   ($[Fe/H] = -2$ to $-1$) as globular clusters in the halos of spiral 
   galaxies, in agreement with \citet{ash93}.
   Turning to the Local Group, \citet{min96} constructed a master-dE galaxy 
   combining the data from
   globular cluster systems of several dEs in this Group; they found 
   an old and metal-poor globular cluster population whose metallicity
   distribution matched the one of the Milky Way halo globulars.     

   The abovementioned bimodality in the color distribution of 
   globular clusters in elliptical galaxies, is closely 
   related to the formation of the globular clusters and a variety of 
   scenarios have been proposed (for reviews on the 
   subject see, e.g.,  \citealt{ash98} or \citealt{har01}). \citet{ash92} 
   and \citet{zep93} predicted 
   this bimodal metallicity distribution of globulars in elliptical 
   galaxies as a result of gas-rich  mergers; they proposed that   
   the blue population originally belonged to the progenitor galaxies 
   and the red population formed during the merger. The numerical 
   simulations from \citet{bek02} showed that dissipative mergers 
   create new globular clusters but they were not able to reproduce
   properly the bimodal metallicity distribution observed in elliptical
   galaxies. 
   
   A different point of view was exposed, e.g.,  by \citet{for97} who found 
   a correlation between the mean metallicity of the metal-rich globular 
   clusters and the parent galaxy luminosity, which suggested that they 
   share the same chemical enrichment process, while the mean metallicity 
   of the metal-poor ones seems to be independent of the galaxy luminosity. 
   They proposed that the bimodality originated in two phases of globular 
   cluster formation from gas with different metallicities, and that most of 
   them formed "in situ". 
   The HST study of 17 early-type galaxies by \citet{lar01} 
   showed a correlation between the colors of both, blue and red globular 
   clusters populations, with the B-luminosity and central velocity dispersion
   of the host galaxy, and concluded that their observations support globular 
   cluster formation "in situ", in the protogalaxy potential well.
   
   Alternatively, \citet{for01} analyzed the relation between the 
   mean color of blue and red globulars with the galaxy velocity 
   dispersion and suggested that red globular clusters share a 
   common origin with the host galaxy and blue ones seem to 
   have formed quite independently; according to \citet{cot98} these 
   blue globular clusters may have been 
   captured from other galaxies by merger processes or tidal stripping.
   
   The idea of the accretion of dwarf galaxies into the cD halo of  
   \object{NGC 1399}, the stripping of their gas and globular clusters 
   and the formation of new clusters from this gas poses a different origin
   for the red globular clusters \citep{hil99}. 
   \citet{bur01} analyzed the blue globular cluster populations from 47
   galaxies and found no correlation between their mean metallicity,
   which is very similar for all these systems, and the galaxy properties 
   (luminosity, velocity dispersion, etc); they proposed that the metal-poor 
   globular clusters may have formed from gas fragments of similar metallicity,
   as already suggested by \citet{ash93}, and located within the dark halo 
   of the galaxy. 
      
   More recently, the semianalytic model by \citet{bea02} assumed that 
   the metal-poor globular clusters formed in protogalactic fragments 
   and the metal-rich ones originated in the gas-rich mergers of such 
   fragments that occurred later. 

   Assuming the presence of globular clusters inside clusters of 
   galaxies, an alternative 
   scenario is proposed by \citet{whi87} and \citet{wes95},
   who pointed to the possible existence of a population of globular clusters 
   that are not bound to individual galaxies; instead, they are supposed to
   move freely in the central regions of the galaxy clusters. These
   intracluster globular clusters may be the result of interactions or mergers
   between the galaxies, or they may have formed precisely in the
   environment of a galaxy cluster without any parent galaxy.
   The kinematic analysis by \citet{min98} and by \citet{kis99} also 
   suggested that some globular 
   clusters may be associated with the gravitational potential of the galaxy 
   cluster and not solely with \object{NGC 1399}. 
   On the other hand, \citet{gri99} found no evidence of intergalactic 
   globular clusters in an HST/WFPC image at a radial distance of about 
   $1\,\fdg4$ from \object{NGC 1399}, but due to the small 
   field of view, they were not able to rule out their existence.
   Several objections against the intraclusters were raised by \citet{har98}
   who tried to explain by means of their existence the supposed high 
   specific frequency of M87, the central Virgo galaxy; but the latest 
   values of 
   $S_\mathrm{N}$ obtained for NGC 1399 by \citet{ost98} and \citet{di02b} 
   showed that it is not so high ($S_\mathrm{N}$ = 5.6 and 5.1, 
   respectively).
    
   In favor of the existence of intergalactic material, \citet{the97},
   \citet{men97}, and \citet{cia98} presented evidence for the presence 
   of intergalactic planetary
   nebulae within the Fornax and Virgo Clusters, while \citet{fer98} 
   reported several hundred of intracluster red giants in Virgo. Some 
   globular clusters may have been stripped with them from other 
   cluster galaxies if this is their origin \citep{har01}.  

   In this paper, we analyze the characteristics of globular
   cluster candidates found near dwarf galaxies in the Fornax Cluster
   and its connection with the above mentioned scenarios.
   It is organized as follows: Section 2 describes
   the observations and the adopted criteria for the globular cluster
   candidates' selection. In Sect. 3 we analyze the color
   distribution, luminosity function and spatial distribution
   of the candidates. Finally, a summary of the results and a
   discussion on their implications are provided in Sect. 4.
   Preliminary results of this work have been presented by 
   \citet{bas02}.
 
\section{Observations and globular cluster selection}

\begin{figure*}
\sidecaption
\includegraphics[width=12cm]{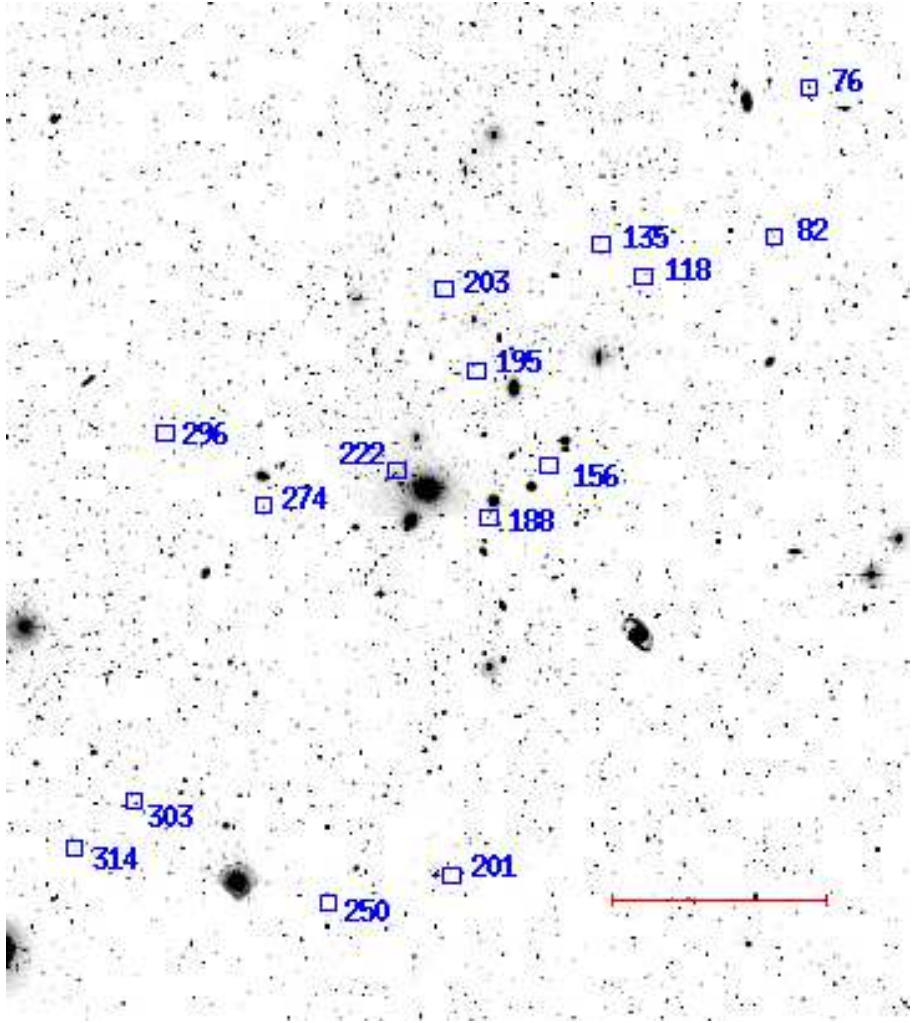}
\caption{DSS image of the Fornax Cluster with NGC\,1399 at the center. 
The squares and the FCC numbers identify the 15 dwarf galaxies. The 
segment corresponds to 1\degr . North is up and East to the left.}  
\label{f1}
\end{figure*}

   CCD images of 15 fields centered on low surface brightness (LSB)
   galaxies in the Fornax Cluster were obtained (with the original purpose 
   of studying the LSB galaxies, \citealt{cel94}; \citealt{cel96}) with 
   the 0.90-m and 1.50-m telescopes at CTIO (Chile),
   during two observing runs in October 1989 and November 1990,
   and using the $C$ and $T_1$ filters of the Washington photometric
   system. The dwarf galaxies are listed in Table~\ref{T1}: the first column 
   gives their FCC numbers \citep{fer89} and the second one gives the 
   respective angular distances from \object{NGC 1399},
   which will be considered as the center of the Cluster. 
   The dwarfs are distributed in 
   angular distance from 12\arcmin~ up to 175\arcmin~ from \object{NGC 1399}
   (see Figure\,\ref{f1}); as we will adopt a distance modulus of 31.35 
   for the Fornax Cluster throughout this paper \citep{mad99}, that 
   corresponds to projected distances ranging between 65 and 
   950 kpc from the cluster center. The fields sizes range from 10 to 
   17 arcmin$^2$; hence, they have the advantage of being larger 
   than the ones studied by \citet{mi98a} to search for cluster candidates. 
   However, they are not 
   as deep as the HST images: we identify cluster candidates up to 
   $T_1 \approx $ 22 mag while Miller et al. reach $V \approx $ 25 mag 
   (equivalent to $T_1 \approx $ 24.5 mag, according to the relation 
   between these magnitudes obtained from \citet{gei96} and \citet{cel94}). 
   For a detailed description of the observations we refer to 
   \citet {cel94}. 

   In order to identify the globular cluster candidates, we selected 
   the point sources
   within certain ranges of colors and magnitudes. The analysis of each
   frame was carried out in the following steps:
   
\begin{enumerate}
      \item  The dwarf galaxies were first removed using the 
   IRAF\footnote{IRAF is
   distributed by the National Optical Astronomy Observatories, which is
   operated by AURA, Inc.\ under contract to the National Science
   Foundation.} / STSDAS tasks ``ellipse" and ``bmodel", which allowed us
   to fit elliptical isophotes to the galaxies, create a model based on 
   them and subtract the model from the image. 

     \item  Using the photometric software SExtractor 
   (Source-Extractor, \citealt{ber96}) we detected all the sources in the
   dwarf-subtracted frames, we measured their instrumental magnitudes
   and colors, and each object was classified with an ``stellarity-index",
   a tool for performing a reliable resolved/unresolved source separation.
   Among the different types of magnitudes that SExtractor measures, the
   ``corrected isophotal" were selected as the most reliable ones
   for stellar-like sources \citep{arn97}. 

      \item The instrumental magnitudes and colors were transformed to the
   standard system via the equations derived by \citet{cel94}, which 
   were applied to the new measurements because they were both obtained by 
   means of aperture photometry. The reddening
   towards the dwarfs' fields was obtained from the \citet{sch98} maps and
   transformed into the Washington system by means of the relations given 
   by \citet{har79}. As the mean estimated color--excess $E_{C-T_{1}} \approx $ 
   0.02 mag is much smaller than the photometric errors (see below), 
   it was considered negligible.

  \begin{table}
     \caption[]{Dwarf galaxies and surface densities of globular 
cluster candidates ($\delta_{GC}$) in their fields.}
        \label{T1}
    $$
        \begin{array}[]{cccc}
           \hline
           \noalign{\smallskip}
     ~~FCC~~ &  d~[arcmin]^{\mathrm{a}}~ &~~\delta_{GC} ~^{\mathrm{b}}~~ &
 ~~\delta_{GC} ~(corrected)~^{\mathrm{c}}~~   \\
           \noalign{\smallskip}
           \hline
           \noalign{\smallskip}
             76 & 174.2 &  0.52 & 0.35      \\
             82 & 139.8 &  0.00 & 0.00      \\
            118 & ~95.7 &  0.56 & 0.39      \\
            135 & ~91.5 &  0.00 & 0.00      \\
            156 & ~42.3 &  0.23 & 0.06      \\
            188 & ~22.7 &  0.46 & 0.29      \\
            195 & ~37.0 &  0.41 & 0.24      \\
            201 & 110.0 &  0.35 & 0.18      \\
            203 & ~56.2 &  0.71 & 0.54      \\
            222 & ~12.1 &  0.58 & 0.41      \\
            250 & 122.2 &  0.00 & 0.00      \\
            274 & ~57.4 &  0.23 & 0.06      \\
            296 & ~92.4 &  0.40 & 0.23      \\
            303 & 135.1 &  0.00 & 0.00      \\
            314 & 159.7 &  0.69 & 0.52      \\
           \noalign{\smallskip}
           \hline
        \end{array}
    $$
\begin{list}{}{}
\item[$^{\mathrm{a}}$] Distance from NGC\,1399.
\item[$^{\mathrm{b}}$] Number of globular cluster candidates/arcmin$^2$ 
found in the corresponding dwarf field, without corrections.
\item[$^{\mathrm{c}}$] Idem $^{\mathrm{b}}$ but background--corrected.
\end{list}
  \end{table}

      \item All the sources classified with an ``stellarity-index" below 0.35
   were considered as resolved sources and were discarded. The limiting
   index was estimated running DAOPHOT on several images as well as adding 
   artificial stars to the frames, and
   comparing the results for different values of the ``stellarity-index" with
   the results for the DAOPHOT indices ``sharpness", ``chi" and ``roundness" 
   that allow the identification of point sources, galaxies and image defects.

      \item Finally, we selected as globular cluster candidates those 
   point sources with:
   magnitudes $19 < T_1 < 22$ mag and colors $0.8 < (C-T_1) < 2.2$ mag.
   We estimate that the completeness of the sample is larger than 90 \%  
   above magnitude $T_1 \approx $ 22 mag. This conservative cutoff 
   prevents against any possible bias introduced by small variations in the 
   photometric limit between different fields. Reassuringly, we found no 
   correlation 
   between effective exposure (considering telescope diameter and integration
   time) and surface density of globular cluster candidates. The number of 
   globular cluster 
   candidates that was identified in each dwarf galaxy's field ranges 
   from zero to ten, which corresponds to surface densities between 0.00
   and 0.71 candidates/arcmin$^2$, as is listed in the third column of 
   Table~\ref{T1}. 
   
\end{enumerate}
  
\begin{figure}
\resizebox{\hsize}{!}{\includegraphics{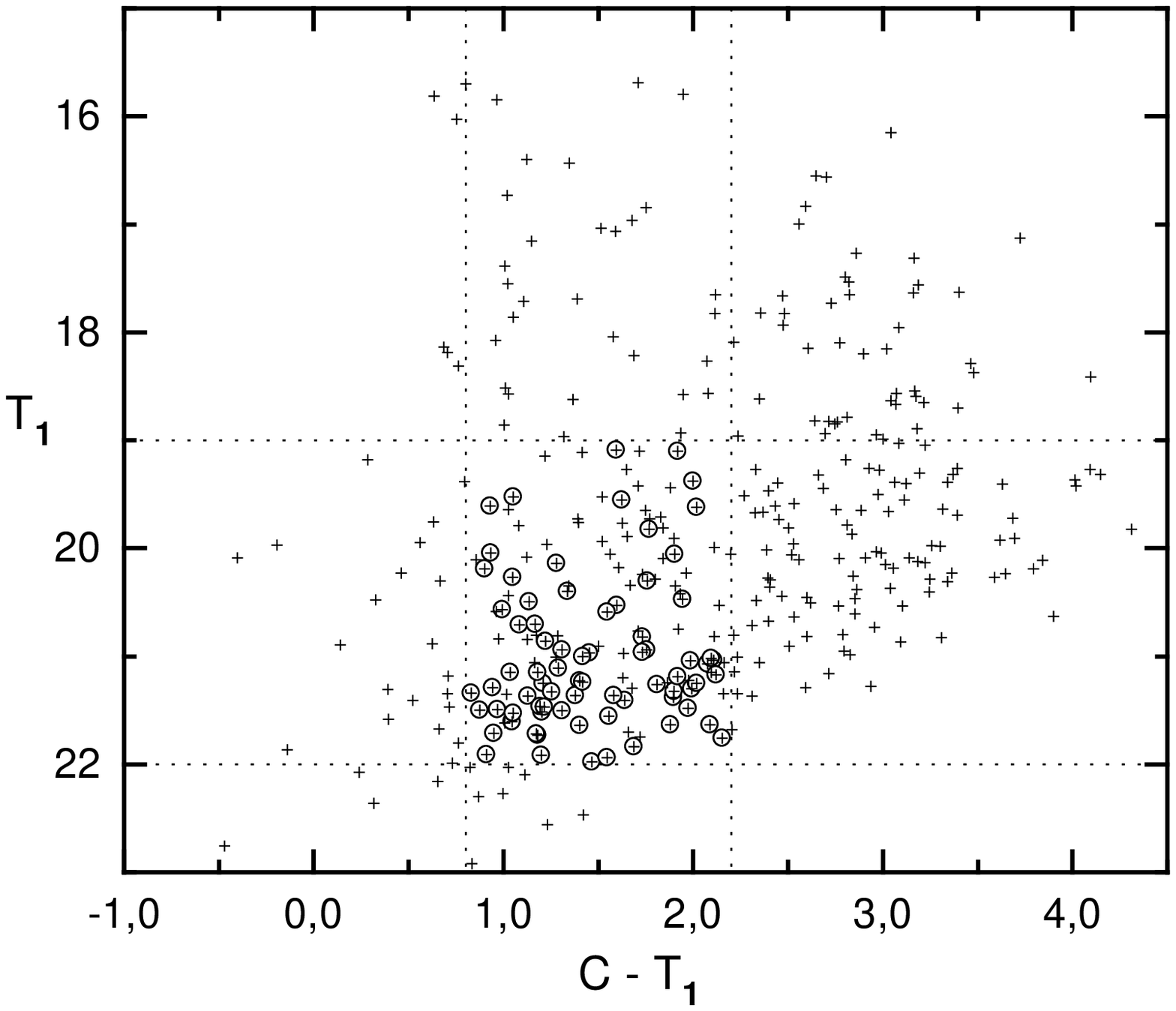}}
\caption{Color-magnitude diagram for all the detected sources (crosses)
and the globular cluster candidates (open circles). Dotted lines represent 
the limits of the globular cluster selection.} 
\label{f2}
\end{figure}

\begin{figure}
\resizebox{\hsize}{!}{\includegraphics{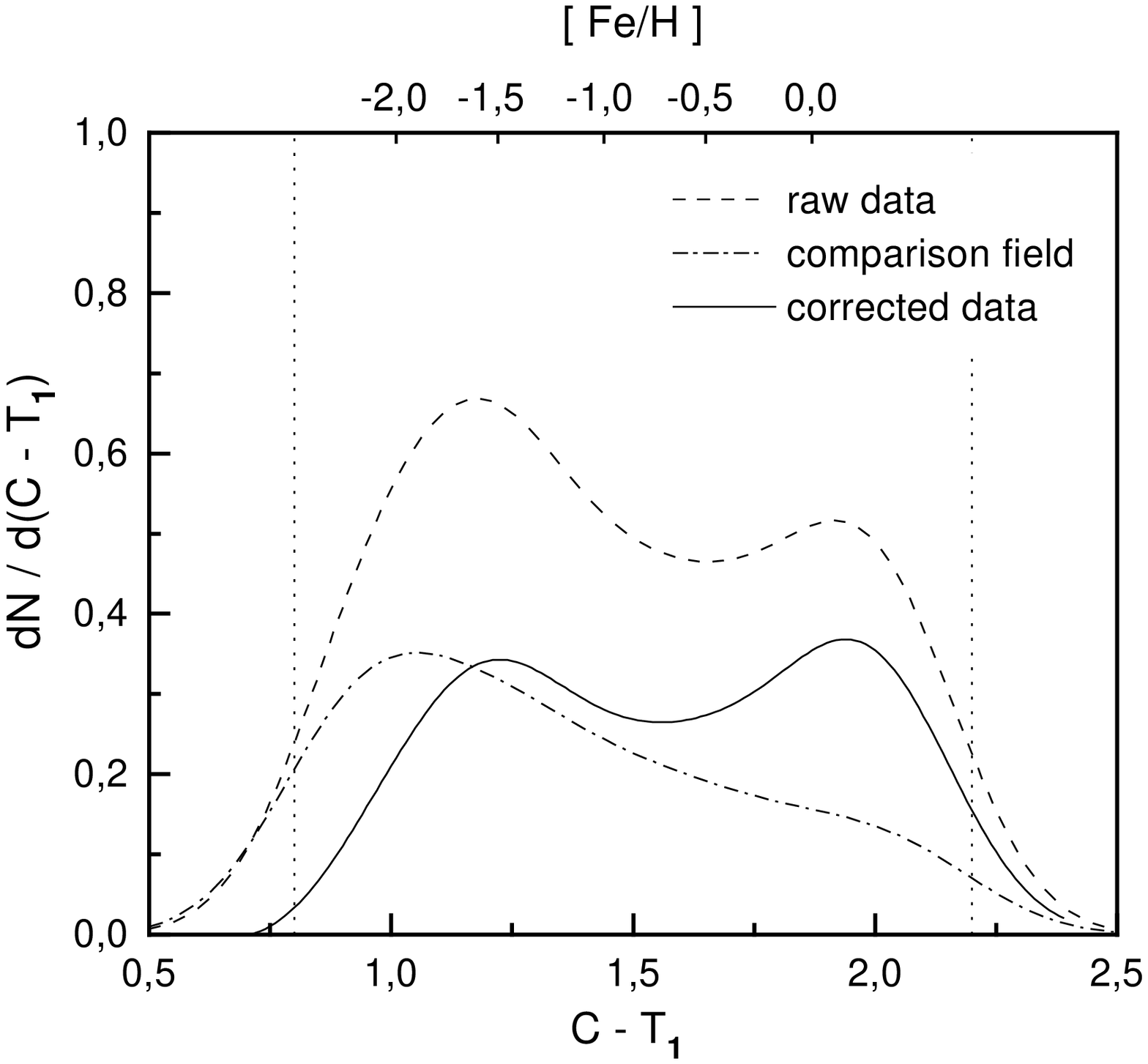}}
\caption{$(C-T_1)$ color distribution: the dashed line corresponds to 
the raw data for the globular cluster candidates, the dot-dashed line 
to the comparison 
field and the solid line shows the result after the background subtraction.
Dotted lines represent the limits of the globular cluster selection.
The metallicity scale \citep{gei90} is given on top.} 
\label{f3}
\end{figure}

    Figure\,\ref{f2} shows the color-magnitude diagram of all the 
   identified sources (about 350), and the ones selected as globular 
   cluster candidates (75).
   The mean photometric errors for the globular cluster candidates are 
   0.09 mag in $T_1$ and 0.15 mag in $(C-T_1)$. 

\section{Results}

   We concentrate on the following features of the globular 
   cluster candidates: 
   their color distribution, their luminosity function, and their spatial
   distribution. Due to the small number of candidates found around each
   dwarf, we will analyze the characteristics of all of them together.

\subsection{\bf Color distribution}

   The $(C-T_1)$ color distribution is shown in Figure\,\ref{f3}; it is 
   an histogram smoothed by means of a Gaussian with a dispersion comparable 
   to the errors in $(C-T_1)$. The color-metallicity calibration from 
   \citet{gei90} has also been included in this Figure.  In order to take
   into account the background contamination we have used a comparison
   field located at $3\,\fdg5$ north-east from the cluster center, which has 
   already been used by \citet{di02b} and whose color distribution is also 
   shown in Figure\,\ref{f3}, after normalizing 
   for the different field sizes corresponding to the dwarfs' images 
   and to the background's one. The comparison field was observed with
   $C$ and $R_{KC}$ filters instead of $C$ and $T_1$, but according to 
   \citet{gei96}, the Kron-Cousins $R$ and the Washington $T_1$ magnitudes 
   are similar, to a high degree of accuracy, for the color range 
   considered in this paper. The comparison field is not very far 
   from \object{FCC 76}, the dwarf with the largest projected distance from 
   \object{NGC 1399} in the sample; thus, the contribution 
   from the background may be probably overestimated and so we may be 
   underestimating the surface densities obtained after this correction.
   The surface density of globular cluster 
   candidates for each dwarf field, after the background-correction, is 
   given in the last column of Table~\ref{T1}. 
   
   The color distribution of the raw data, as can be seen in 
   Figure\,\ref{f3}, is extended and appears to be bimodal 
   (see also the color-magnitude diagram displayed in Figure\,\ref{f2}). 
   In order to quantify this apparent bimodality in a statistical way, 
   we have applied to the raw data the KMM test which helps to detect 
   and evaluate 
   bimodality in datasets (see \citealt{ash94} for a description of the 
   test and its application). The results of the test indicate 
   that two Gaussians with means at 1.18 and 1.86 mag can be fit to 
   the set of $(C-T_1)$ values, and assuming that both Gaussians have the 
   same covariance (homoscedastic fitting) we obtain a dispersion 
   $\sigma$~=~0.2 mag for them. The hypothesis that this  $(C-T_1)$ 
   distribution is unimodal rather than bimodal is rejected at a confidence
   level of 100\% according to the KMM test. It must be taken into 
   account that the number of candidates is small (75 objects) but, according 
   to the study of the KMM algorithm sensitivity performed by 
   \citet{ash94}, there is a sufficient large separation in the means of
   the two Gaussians (3.4~$\sigma$) to be able to obtain a 
   significant rejection of the unimodal hypothesis. 
   
   The result of the background subtraction is also shown in Figure\,\ref{f3}, 
   where the corrected color distribution appears to be bimodal too, with 
   two possible peaks that would  be located at $(C-T_1) \approx$ 1.2  
   and 1.9 mag, 
   corresponding to metallicities [Fe/H] $\approx  -1.6$  and 0.1, 
   respectively \citep{gei90}. 
   However, it is not possible to apply the KMM test to this corrected 
   distribution because the 
   number of candidates is smaller than 50, and \citet{ash94} state that 
   in this cases the test does not provide a reliable result for 
   detecting bimodality. Anyway, even though we cannot confirm statistically 
   the bimodality, 
   it is clear that we do not find a single population of metal-poor
   globular cluster candidates around the dwarfs, as happens in the
   cases already mentioned in the Introduction; instead, an extended 
   distribution that expands over the whole metallicity range, from
   metal-poor to metal-rich populations, is detected.

   In spite of the apparent symmetry in the corrected color distribution
   displayed in Figure\,\ref{f3}, we cannot be sure that the number of 
   these probable metal-poor and metal-rich candidates are similar 
   because we do not
   have a complete area sample. Figure\,\ref{f1} shows that there
   are more dwarfs near to \object{NGC 1399} than far from it. 
   We have not attempted to search for radial globular clusters 
   color gradients with respect to the dwarfs or to the Fornax Cluster 
   center due to the small sample we are considering.  
  
   Whether or not two globular cluster populations should be expected in
   dwarf galaxies is still unclear. 
   All the globular clusters around dwarfs in the Local 
   Group studied by \citet{min96} had metallicities that correspond to
   a metal-poor population, with [Fe/H] $\leq  -1$. They suggested
   that the dE galaxies included in their study seem to have formed no 
   metal-rich globular clusters. In turn, \citet{du96b} found a possible 
   bimodality in the colors of 
   the globular clusters of the Virgo dwarf VCC 1254, which they 
   speculate might correspond to two phases of globular cluster formation. 

   With regard to the results for the \object{NGC 1399} globular 
   cluster system, \citet{ost98} identified  globular clusters 
   located between $0\farcm5$ and $4\arcmin$ from the galaxy center 
   finding two globular cluster populations with 
   $(C-T_1) \approx$ 1.3 and 1.8\,mag, respectively. In turn, 
   \citet{di02b} performed a wide-field study of this system 
   and showed that the innermost sample, between 1\farcm8 and 4\farcm5, 
   was clearly bimodal with peaks at $(C-T_1) \approx $ 1.3  and 1.75\,mag; 
   at larger radii, from 4\farcm5 up to $22\arcmin$, the metal-poor population 
   became dominant without changing the position of the blue peak. 

\subsection{\bf Luminosity function}

\begin{figure}
\resizebox{\hsize}{!}{\includegraphics{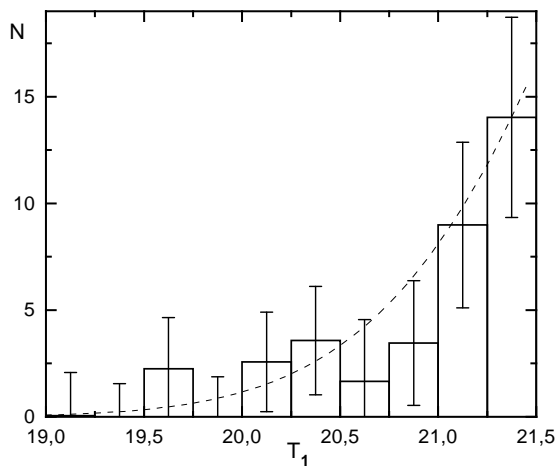}}
\caption{Luminosity function (background-corrected) for the globular cluster 
candidates. The dashed line represents a Gaussian with 
$< T_1 > = 23.3$ mag and $\sigma~=~1.2 $ mag. Errors are based on 
Poisson statistics.} 
\label{f4}
\end{figure}

   We plot the luminosity function (background-corrected) in Figure\,\ref{f4}: 
   an histogram of the number of globular cluster candidates vs.\ $T_1$, 
   where we use an upper limiting magnitude, $T_1 = 21.5$ mag, omitting 
   the last bin where the
   completeness of the red candidates is more seriously affected. 
   A Gaussian distribution, calculated with the parameters obtained 
   by \citet{ost98} fitting the luminosity function of the \object{NGC 1399} 
   globular cluster system, is included for comparison.
 
   Our sample covers 
   only $7\,\%$ of the area under that Gaussian and, after background 
   subtraction, we are left with 37 globular cluster candidates with magnitudes
   $19 < T_1 < 21.5$. Although it is a small sample, we attempt to 
   compare it with the number of globular clusters that we should have found, 
   in a similar area, if they belonged to the dwarf galaxies' globular 
   cluster systems.   
   We calculate the total integrated brightness of our sample's dEs by
   means of the $T_1$ total integrated magnitudes of the dwarfs given by 
   \citet{cel94}, the adopted distance modulus to the Fornax Cluster and 
   the relation between V and $T_1$ magnitudes mentioned above; thus, we 
   obtain a total integrated visual brightness for the sample of $-18.8$ mag. 
   As the range of specific frequencies proposed for dEs is 
   $S_\mathrm{N}$ = 2--6 \citep{mi98a, elm99}, we estimate that we 
   should have found between 4 and 13 globular cluster dwarfs' GC candidates, 
   within the mentioned $T_1$ range. By comparison, we have identified 
   three to ten times more globular cluster candidates than what is inferred 
   from the $S_\mathrm{N}$ values. 

\subsection{\bf Spatial distribution}

\subsubsection{Distribution with respect to the dwarf centers}

   If the globular cluster candidates are bound to the respective 
   dwarf galaxies, the projected density of globular clusters vs.\ 
   galactocentric distance is expected to increase towards the center. 
   This behavior can be clearly seen in the Fig.\,5 of \citet{lot01}, 
   which shows the summed radial distribution of globular cluster 
   candidates from a sample of 51 dEs.
   
\begin{figure}
\resizebox{\hsize}{!}{\includegraphics{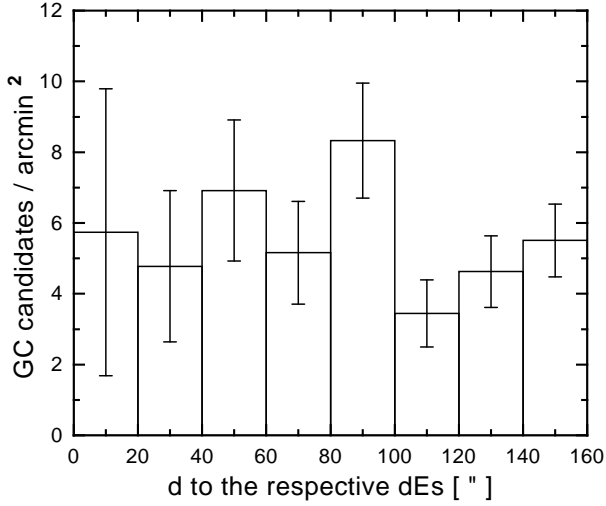}}
\caption{Surface density distribution of the globular cluster candidates 
with respect to the dwarf galaxies. Errors are based on Poisson statistics.} 
\label{f5}
\end{figure}

   Figure\,\ref{f5} depicts the surface number 
   density of globular clusters (background-corrected), calculated in 
   concentric annuli 
   around each of the dwarfs and summed over all of them.
   These surface densities are estimated as follows. First, 
   a set of 20\arcsec wide annuli is established around each dwarf, 
   taking into account the different scales of the images.
   For each image, annuli with more than 60\% of their area outside
   the frame limits were discarded, thus leading to the 160\arcsec~ 
   limit in angular distance. 
   Then, for each dwarf, the number of globular candidates is estimated
   within each annulus, applying a completeness correction to the annuli
   lying partly outside the corresponding frame. Afterwards, the counts 
   for each annulus are 
   background-corrected, subtracting the density of the background field 
   multiplied by the area of the corresponding complete annulus. Finally, 
   the background-corrected counts are summed over the annuli defined by 
   the same angular distance from each dwarf, and divided by its complete 
   area. 
   
   As can be seen in Figure\,\ref{f5}, no concentration towards the center 
   is evident.
   In order to demonstrate this statistically, we compare this observed
   distribution with a uniform distribution, that is, one with a constant 
   number density calculated as the same number of globular clusters, 
   scattered 
   across the same total area. The result of a $\chi ^2$ test performed 
   between them, indicates that the observed distribution is statistically 
   consistent with being drawn from a uniform distribution at a significance
   level of 89\,\%. Under this evidence, it is not possible to assert 
   that these globular cluster candidates are bound to the respective dwarfs
   as they show no concentration to the dwarf centers, 
   although we cannot confirm this hypothesis without the aid of radial 
   velocities. 
  
    We must 
   take into account that the studied radial distribution shown in 
   Figure\,\ref{f5} extends up to 14 kpc from the center of the dEs. 
   The radial density profiles from \citet{lot01} reach almost zero  
   value at shorter distances from the dwarfs, between 1.1 and 
   8.7 kpc, according to the dwarf's exponential scalelength they use
   and the distance modulus we have adopted, 
   while the summed radial distribution of the globular cluster system for 
   11 Virgo dEs
   by \citet{du96b} extends to only 2.5 kpc. Both results indicate that the 
   globular cluster systems of dwarfs are rather compact.

\begin{figure}
\resizebox{\hsize}{!}{\includegraphics{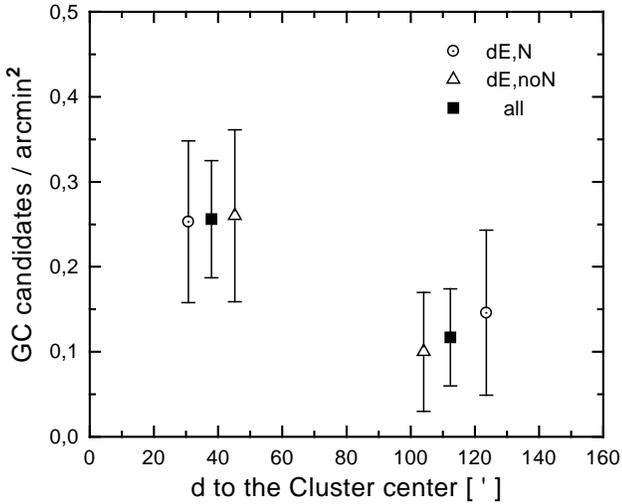}}
\caption{Surface density of the globular cluster candidates 
(background-corrected) with respect to the Fornax Cluster center 
(\object{NGC 1399}) grouped according to its angular distance from it 
(less than or greater than 80\arcmin). 
Dotted circles correspond to the nucleated dEs, open triangles to 
the non-nucleated dEs and filled squares to all the dE galaxies. 
Errors are based on Poisson statistics.}  
\label{f6}
\end{figure}

\subsubsection{Distribution with respect to the Cluster center}
   
   Finally, we studied the surface density of the potential globular 
   clusters but 
   now vs.\ the distance to the center of the Fornax Cluster, more precisely 
   to \object{NGC 1399} (Figure\,\ref{f6}). As the number of GC candidates 
   is small,  we consider them in two separate groups to reinforce 
   the statistic: the globular clusters located nearer than 80\arcmin~ 
   from \object{NGC 1399} 
   and those between 80\arcmin~ and 160\arcmin~ from it. 
   
   The surface densities of globulars were calculated as the number of 
   clusters 
   (background-corrected) around the dwarfs corresponding to each group, 
   divided by the total surveyed area in each case.   
   We have excluded the two dwarf galaxies more distant from the center, 
   at more than $2\,\fdg5$: that is \object{FCC 76} and \object{FCC 314}. 
   \object{FCC 76} has 
   probably underwent recent star-formation \citep{cel94} while \object{FCC 314}
   is seen projected on to a background cluster of galaxies, so both of them 
   could led to misleading results for the  distribution of globular 
   cluster candidates with respect to the cluster center. 
   
   Although the statistical noise is relatively large, the results show 
   that the globular clusters located 
   closer to the cluster center have a higher projected density than the ones 
   located farther than 80\arcmin, which seem to be approaching to zero 
   density. As the $S_\mathrm{N}$ values for nucleated dEs are higher than 
   for non-nucleated dEs \citep{mi98a}, we also estimate separately the 
   surface density of the globular cluster candidates around dwarfs with 
   nucleus and without it. They are included in Figure\,\ref{f6} and show 
   no difference between them or with the results for all the dEs together.

\section{Discussion}

   With regard to the color distribution of the globular cluster 
   candidates, it is 
   interesting to note that we do not find a single population, the 
   metal-poor one, as seems to be the common case for dwarf galaxies 
   \citep{du96b,mi98a,mi98b,mil99},  but an extended distribution, 
   which appears to be bimodal though we cannot prove it statistically due
   to the small sample involved. In addition, if we take into account 
   the specific 
   frequency estimated for dwarf galaxies \citep{mi98a,elm99} and the 
   luminosity function of the globular cluster candidates, we should have 
   found significantly fewer globular cluster candidates than we actually do. 

   According to the projected density of the potential globular clusters, 
   they show no concentration towards the dwarfs while they do show 
   concentration with respect to the center of the cluster. These results 
   led us to speculate that the globular cluster candidates may not be 
   associated to the dwarf galaxies themselves. We are then left with
   three possibilities: first, that these globular cluster candidates 
   belong to the globular cluster system
   of \object{NGC 1399}; second, they may be moving freely throughout 
   the potential well of the cluster, without being bound to
   any galaxy in particular; or third, that they are a mix of both.                             
   In the first case, we should be accepting that the globular cluster 
   system of \object{NGC 1399} 
   is much more extended and numerous than previously thought. According to 
   this hypothesis, Figure\,\ref{f6} suggests that there should be clusters 
   up to at least an intermediate angular distance of 80\arcmin~(a projected 
   distance of about 430 kpc) from the central galaxy; the CCD study over 
   the largest area around \object{NGC 1399} was performed by \citet{di02b},
   which extends up to 22\arcmin, and showed that the globular cluster 
   system extends over a radial projected distance of more than 100 kpc.

   The total number (background corrected) of globular clusters within
   the area covered by our observations can be roughly estimated as about
   550 clusters, by extrapolating their luminosity function over the
   whole range of $T_1$ magnitudes. Thus, the number of clusters
   that should be distributed within a circular area of radius 80\arcmin
   around \object{NGC 1399} may be calculated, just taking into account
   the ratio of the areas, as several $10^4$ clusters. For
   comparison, the total number of globular clusters associated with 
   galaxies in the Fornax Cluster may also be roughly estimated as
   follows. The blue magnitude of all the galaxies included in the
   \citet{fer89} catalogue is $B = 8.3$ mag; adopting for them a mean
   color index $B-V \approx 0.8$ mag and the distance modulus mentioned
   above, we obtain an absolute visual magnitude $M_V = -23.8$. If we
   assume a "typical" specific frequency $S_\mathrm{N}=5$ we conclude
   that about $1.6\times 10^4$ globular clusters should be associated
   with galaxies in Fornax.  It is interesting to note that the number of
   globular clusters that we found within a circular area of radius
   80\arcmin around \object{NGC 1399} is of the same order o larger than
   the estimated number of globulars associated to galaxies in the whole
   Fornax Cluster.

   It is also likely that some globular clusters might 
   have escaped 
   from its parent galaxies and, after that, remained within the 
   potential well of the Fornax Cluster as a whole \citep[see, for 
   instance,][]{kis99}.  \citet{whi87} propose that the distribution 
   of the stripped globular clusters within the cluster
   will follow the same density profile as the galaxies and they might 
   form a kind of envelope around the central galaxy. Alternative 
   origins for the intraclusters are mentioned by \citet{wes95}, 
   who speculate that they might have formed ``in situ", without a parent
   galaxy, or during mergers of sub-systems with a high gas content. 

   Deeper images are required to clarify this picture and a new survey in 
   the Fornax Cluster is in progress. The true nature of these candidates
   might be confirmed by means of spectra.

\begin{acknowledgements}

   We wish to thank the referee, Dr. M. Kissler-Patig, for his comments which
   helped to improve the present paper. We are also grateful to S. D. Abal,
   M. C. Fanjul and R. E. Mart\'{\i}nez for technical assistance. This work 
   was partially supported by grants from CONICET and Fundaci\'on Antorchas, 
   Argentina.

\end{acknowledgements}

\end{document}